\documentclass[12pt, a4paper]{article}
\usepackage[dvips]{graphicx}

\begin{document}
\date{}
\sloppy
\begin{center}
{\LARGE \bf Detection of Slow Magnetic Monopole}
\end{center}

\begin{center}

{\large \bf Ianovski V.V.~$^1$, Kolokolov I.V.~$^2$, Vorob'ev P.V.~$^2$  \\}

\vspace{1cm}
{\it $^1$ PNPI IANOVSKI@pnpi.spb.ru}\\
\vspace{0.5cm}
{\it $^2$ Budker Institute of Nuclear Physics (BINP)\\
RU-630090 Novosibirsk Academician Lavrentiev
Prospect 11 Russia.\\
P.V.VOROBYOV@inp.nsk.su, I.V.Kolokolov@inp.nsk.su }
\end{center}

\begin{abstract}
{\it  Numerous and  all  unsuccessful  attempts  of  experimental
search for monopole in cosmic rays and on  accelerators  in  high
energy particle collisions have been done since the possibility
of existence of a magnetic monopole has been
surveyed in 1931. Also the searches have been carried out in mica
for monopole tracks as well  as for  relict  monopoles, entrapped
by ferromagnetic inclusions in iron-ores, moon rock and meteorites.

A new method  of  search  for  supermassive cosmic and relict
monopoles  by
magnetically ordered film is considered. This approach  resembles
the traditional method of nuclear  emulsion  chamber.  Apparently
the proposed method is particularly attractive for  detection  of
relict monopoles, released from melting iron ore.}
\end{abstract}

\vspace{1cm}
This work was presented  at 10th International Baksan
School "Particles and Cosmology", April 1999, Baksan Valley

\vspace{4cm}
{\large (C) PNPI, 1999}

Preprint PNPI EP-38-1999 N 2323.

\newpage

\tableofcontents

\section{Introduction}

A concept of a magnetic monopole has been introduced into  modern
physics in 1931 by Paul Dirac \cite{Dir}. He postulated existence
of an isolated magnetic charge $g$. Using general  principles  of
quantum mechanics, he  has  related  the  electric  and  magnetic
charge  values:  $ge=\frac{n}{2}  \hbar  c$,  where  $e$  is  the
electron electric charge, $\hbar$ is the Plank constant,  $c$  is
the speed of light, $n= \pm 1,2...$ is an integer. Numerous  but
unsuccessful  attempts of experimental search for these magnetic
monopoles at accelerators and in cosmic rays   have  been  done
since then\cite{Groom,Klapdor,Barkov,D0}.

The new interest  to  this  problem  has  arised  in  1974,  when
Polyakov \cite{Pol} and 't~Hooft \cite{tHooft}  have  shown  that
such objects exist as solutions in a wide class  of  models  with
spontaneously broken symmetry .

The magnetic charge of  the  Polyakov---'t~Hooft  monopole  is  a
multiple of the Dirac one $g=2 ne/\alpha$, and the value  of  its
mass $M_{g}$ lies in the range of $10^8$ --- $10^{16}$ GeV. Such
GUT's monopoles are not point-like and have complex structure

It is completely clear, that the Polyakov---'t~Hooft massive  monopoles
can not emerge at accelerators,  therefore  we  do  not  consider
accelerator experiments. Moreover, we assume that  the  monopoles
that reach the surface of the Earth are gravitationally bound  up
either with the Galaxy (if $\beta=v/c< 10^{-3}$) or with the  Sun
(when $\beta<10^{-4}$).

The monopole ionization loss in matter has been evaluated by many
authors (look at the reviews: \cite {Groom, Klapdor, Mono}).  For
fast monopole the ionization loss appreciably (about 4700 times!)
exceeds the loss for the minimum  ionizing  particles  ---  MIPs,
which is $dE/dl\simeq 2~MeV/g$.

In units of the ionizing loss of  particle  with  charge  e,  the
monopole ionization loss is given by:
\begin{equation}
\left(\frac{dE}{dl}\right)_g
=\left (\frac{dE}{dl}\right)_{e}
\left (\frac{g}{e} \right)^2 \beta^2 ~.
\label{lost}
\end{equation}

If we recollect that ionization loss of  a  charged  particle  is
proportional to $1/ \beta^2 $, then it is clear, that the loss of
a monopole does not depend on velocity. It should be pointed  out
here that in GUT the monopole is  a  very  heavy  particle.  {\bf
Therefore  supermassive   monopole    is    practically    always
non-relativistic!}

When $\beta \sim 10^{-3}$  the  ionization  loss  of  a  monopole
decreases to level of energy loss of MIP. At $\beta<10^{-4}$  the
ionization  mechanism  of  the  energy  loss  is  switched    off
practically, because in this case  the  energy  of  monopole-atom
collision already is not enough for ionization of the latter.

To estimate the maximum of monopole velocity $v$, it  is  natural
to take the velocity of the  Sun  relatively  to  the  background
radiation
\begin{equation}
\beta = \frac {v} {c} \simeq 10^{-3}~.
\label{vc}
\end{equation}
We shall remind here, that the virial  velocity  for  our  Galaxy
does not exceed $10^{-3}c$ too.

For detection of a slow monopole with efficiency close to 1,  the
superconducting  induction   detectors    were    designed    and
constructed\cite{Cabr}. The single event, registered by the Cabrera
detector, has originated the burst of experimental activity of searches
for monopoles of cosmic origin. However, the significant progress
in sensitivity (and corresponding limits on the monopole  flux)  has
been achieved only  for  ionizing  detectors  \cite{Groom,MACRO}.
Sensitivity of these detectors  is  close  to  extended Parker
bound \cite{Adams}:
\begin{center}
$\cal F \leq$ $1 \cdot 10^{-16}(m/10^{17}~GeV)~cm^{-2} s^{-1}~sr^{-1}.$
\end{center}

One of the primary aims of the MACRO detector at the Gran Sasso
underground lab (in Italy, at an average depth of 3700 hg/cm$^2$)
is the search for magnetic monopoles for a large range of
velocities at a sensitivity level well below the Parker bound for
a large range of velocities, $4 \cdot 10^{-5}<\beta<1$,
$\beta=v/c$. MACRO uses three different types of detectors: liquid
scintillators, limited streamer tubes and nuclear track detectors
(CR39 and Lexan) arranged in a modular structure of six
``supermodules". They consider magnetic monopole with enough
kinetic energy to traverse the Earth $4 \cdot 10^{-5}<\beta<1$,
$\beta=v/c$. \cite{macro_l}

Another possibility of detection of a slow  GUT's monopole is  the
search for proton decay induced by a heavy  monopole  \cite{Rub},
\cite{Cal}. Recently the  group  working  with  the  Baikal  lake
Cherenkov detector \cite{Belo} has set the following limit on the
flux of heavy magnetic monopoles and the Q-balls, which are  able
to induce the proton decay
\begin{center}
$\cal F$ $< 3.9 \cdot 10^{-16} cm^{-2} s^{-1} sr^{-1}.$
\end{center}

Here  we  consider a possibility of building-up of a
new type  of  detector of  slow  monopoles.
Our  idea  is  based  on  registration   of
interaction of a slow cosmic-ray-related monopole with  the  film
of easy-axis and high coercitivity ferromagnet.  As  a  sensitive
element of such a detector one can use an advanced
storage media, namely  --  the  magneto-optical  disk  (MO  disk)
\cite{MO}. (To our knowledge -- for  modern  MO  disks  an  areal
density of $45~Gbit/in^2$ has been demonstrated using  near-field
techniques,  with  a  theoretical  possibility  in   excess    of
$100~Gbit/in^2$). The slow monopole which is passing through  the
magnetic coat of MO disk, leaves a  distinctive  magnetic  track,
and this track can  be  detected. It is important to note that
considerable surface  can
be covered by  MO disks. They can be exposed  any  reasonable
time  without  any  maintenance,  like  in    emulsion    chamber
experiments or  the  CR39  nuclear  track  subdetector  of  MACRO
\cite{CR39}.

Apparently, the use of such passive detectors will be  especially
effective  in  search  of  the  relict  monopoles,  entrapped  in
ferromagnetic inclusions of iron ore. Such  monopoles  should  be
extracted  from  the  ore  during  the  melting  process.   These
monopoles are extracted at relatively small cross-section of  the
furnace and freely fall downward. Such slow moving monopoles can
be detected by a passive detector with MO disks. These disks  are
to be placed, e.g. in a cavity under the  furnace.  The  effective
exposition time, normalized to the mass of ore,  from  which  the
monopoles are extracted, can be very large. We  wish  note,  that
during the exposure no  detector  service  is  required.  After
exposure the MO disks should further be placed into specialized
device to find the traces of the magnetic monopole.
Ours real, but for the future aim consists in study of a
possibility of making of such specialized device.

\section{Track formation by the slowly moving monopole.}

It is expected that a slow monopole, moving transversely  through
a magnetized ferromagnetic film, should leave a distinctive track
of magnetization in it. We can use this phenomenon to  design  an
effective detector of supermassive  cosmic  monopoles.  For  this
purpose we shall consider in detail  the  mechanism  of  magnetic
track formation by such a type of monopoles \cite{VK,e-print}.

Let us consider a thin  layer  of  easy-axis  hard  ferromagnetic
magnetized perpendicularly to the surface along an easy axis.  It
is easy to see that the external magnetic field is absent (double
layer!), but the surface density of  a  magnetostatic  energy  of
such configuration is rather large. However,  if  the  anisotropy
constant $K_u$ is reasonably large, and the  effective  field  of
anisotropy exceeds the value  of  demagnetizing  field  then  the
system is in a metastable state. As film magnetization is inversed
in a small area, a stable cylindrical domain (magnetic bubble)
of  certain radius is created. Such one domain state is
characterized by an equilibrium radius $r_{eq}$ and a
collapse radius $r_c$ where $r_{eq} > r_c$.
As the domain radius is decreasing and approaching to the
collapse one, $r_c$, the magnetic bubble becomes unstable and
disappears.  A typical magnetic bubble size is about
30 nm for film thickness of 100 nm.

What is the field of a monopole at such distance?
\begin{equation}
H= \frac{\Phi_0} {4\pi r_0^2} \simeq 2\cdot 10^3~Oe~,
\label{bmon}
\end{equation}
 that is enough without any doubt for re-magnetization of a
material with coercitivity of the order of 1~Oe. We have introduced
here some characteristic radius $$r_0 = 2\mu_0 \gamma / I_s^2$$
which defines the  minimal  radius  of collapse.

All above-mentioned is true for films with high mobility of domain
walls. Films with low wall mobility are even more stable, and  at
the same time, domains with radius less than the collapse  radius
can exist in them, in principle. So, in  the  $Co/Pt$  films  the
movement of domain walls is suppressed. And in the 20 nm film the
transverse domains of cylindrical form with diameter of the order
of 50-100 nm are obtained. Let us note, that the coercitivity
of the  easy-axis  $Co/Pt$  film  is  of  the  order  of  1-2~kOe
\cite{Mopt}.

However, our speculations are true only in static, for very  slow
monopoles only. As  we  have  noted  before,  the  characteristic
velocity of a monopole is $v \simeq  10^{-4}-10^{-3}c$,  and  for
our  consideration  let  us  assume  $v=10^{-4}c$.  The  time  of
monopole interaction with an electron $\tau$ can  be  defined  as
the time, during which a field higher than  some  critical  field
$H_c$ interacts with the electron
\begin{equation}
\tau \simeq \frac{r_c}{v} \simeq \frac{1} {v} \sqrt {\frac {\Phi_0}
{4\pi H_c}} ~.
\label{tmon}
\end{equation}

At $H_c$ of the order $3\cdot10^3$ Oe we have
$\tau \sim 3\cdot 10^{-12} ~s.$
It means, that the spin flip of the magnetic in the "track" takes
place during the interaction time.

For such spin flip, the adiabatic condition is necessary ---  the
frequency of spin precession in the overturning field  should  be
much larger than the inverse time of the interaction
\begin{equation}
\omega = \frac {\mu_B H} {\hbar} \gg \frac{1}{\tau}~.
\label{wmon}
\end{equation}

It is possible to derive from here  the  minimal  magnetic  field
which is appropriate for the adiabatic mode, and the track radius:
\begin{equation}
H \gg H_c = \frac{\hbar}{\mu_B \tau} = \frac{4\pi\hbar^2 v^2}
{\mu_b^2 \Phi_0}~,
\label{Hc}
\end{equation}

\begin{equation}
R_t \ll r_c = \sqrt {\frac{\Phi_0} {4\pi H_c}}~.
\label{rt}
\end{equation}

In our case at $v \simeq 10^{-4}c$ we get $$H_c \simeq
10^{7}Oe~;~~r_c \simeq 10^{-7}cm,$$
and for $v \simeq 10^{-6}c,$ we have
$$H_c \simeq 10^{3}Oe~;~~ r_c \simeq 10^{-5}cm.$$

It is obvious, that the conditions of adiabatic and even resonant
spin flip are not fulfilled, while $r_c < r_0$, that  corresponds
to the monopole speed $v \simeq 10^{-6} c$.

We shall consider the influence of the conductivity of  the  film
material now. The reason is that the monopole  magnetic  flux  is
being frozen into the cylindrical area around the track axis  and
then spreads radially. The  radius  of  the  flux  pipe  and  the
diffusion  factor  of  the  flux  are  determined  by  the   film
conductivity. The radius of the  flux  pipe  $\delta_c$,  we  can
find:
\begin{equation}
\delta_c =\frac{c^2} {2\pi\cdot \sigma\mu v} ~.
\label{skinc1}
\end{equation}

This length has a simple physical meaning. At distances less than
$\delta_c$ the monopole field  can  be  considered  as  free.  At
distances of the order of $\delta_c$ and more, the magnetic  tail
of the monopole is formed, which is an analogue of a string in  a
superconductor. Due to the finite conductivity of  material,  the
tail spreads gradually, and the  energy  of  the  magnetic  field
converts into heat.

At the monopole velocity about $v \simeq 10^{-4}c$, the flux pipe
has the radius of the order of $10^{-5}cm$. The flux pipe spreads
in time as:

\begin{equation}
R(t) = \delta_c \sqrt {\frac{t}{\tau}}~.
\label{rfromt}
\end{equation}

Thus the magnetic moment of the track is conserved,  as  well  as
the frozen flux. It is easy to calculate the average intensity of
the magnetic field in the flux pipe  immediately  after  monopole
flight

\begin{equation}
H = \frac{\Phi_0} {\pi \delta_c^2} \sim 10^{3} Oe ~.
\label{Hpip}
\end{equation}

The typical time of the monopole interaction with an electron  in
a conductor is:

\begin{equation}
\tau \simeq \frac {\delta_c}{v} \simeq 10^{-11}-10^{-10}~s~,
\label{timpip}
\end{equation}
and the field strength, providing the adiabatic inversion of  the
magnetic spin in a track, will be:

\begin{equation}
H_c \simeq \frac {\hbar} {\mu_B \tau } \simeq 10^{4}~Oe~.
\label{Hcrit1}
\end{equation}

Thus,  the  frozen  field  in  the  conductor  $H_c$  can  effect
appreciably the process of spin flip in the track and provide the
adiabatic spin flip of electrons  in  the  magnetic  at  monopole
speeds below  $10^{-4}~c$.

\section{Detection of the monopole track.}

As it was shown, the domain induced by  the  moving
monopole  has  the  typical  size  of  the  order  of  50~nm  and
magnetization about  several  thousand  Gauss.  Then  the  domain
magnetic flux $\Phi_d$ will be of the order:
$\Phi_d=\pi r^2\cdot I_s\simeq \Phi_0=2\cdot
10^{-7}~G\cdot cm^2$

For detection of such a  flux  we  can  use  the  high  sensitive
fluxmeter on the basis of  superconducting  quantum  interference
device --- SQUID, or magneto-optical device on the basis of  Kerr
effect (rotation of the polarization plane of light reflected  by
a surface of a ferromagnet which is magnetized  perpendicular  to
the surface). It is clear that  the  second  way  is  technically
easier and does not require a cryogenic maintenance. In the later
case, the realization of  such  a  detector  requires  a  surface
covered with a thin layer of easy-axis magnetized magnetic media,
plus a magnetooptic device to  detect  the  spots  of  transverse
magnetization of the film (to detect  the  domain  of  opposite
direction of magnetization!) with a system of precise positioning.

A similar technique has  emerged  recently  in  an  almost  ready
shape, suitable for the detector design with minimal  adjustment.
It is the  magneto-optic  recording  technology  used  in  modern
magneto-optic disks (MO disks) and their readout devices. Already
there are  MO  disks  with  multilayer  coating  of  $Sm/Co$  and
$Pt/Co$. The coercitivity of multilayer coats  $Pt/Co$  is  about
$1~KOe$    at    10    layers    of    total    thickness    about
$15~nm$~\cite{Mopt}. The size of a magnetic bubble which can  be
detected in such a coat by the magneto-optical  method  is  about
$60~nm$. This technique using the  near-field  optics  has  been
designed, f.i. in Bell Laboratories \cite{Bell}.

The coercitivity  of  coats  with  $Sm/Co$  is  in  the  interval
$3-5~kOe$,  and  the  reference  size  of  magnetic  bubble    is
$50~nm$~\cite{Mopt}. For detection of the magnetic track of  the
monopole  it  is  possible  to  use  slightly  modified  standard
MO-drives. Having  covered  a  large enough surface with such MO
disks, we can obtain an effective and
relatively nonexpensive detector of slow moving space  monopoles,
which we can expose during unlimited time.

However, it is more effective to use the  MO-detector  to  search
for relict monopoles, entrapped in  ferromagnetic  inclusions  of
iron ore. Naturally, the melted iron ore becomes paramagnetic and
the ferromagnetic traps disappears. Then the monopole is likely to
be surrounded by a cluster of several dozens of iron  atoms.  The
size of a complex is  determined  by  thermodynamic  equilibrium:
\begin{equation}
\frac{\mu_{Fe}\cdot g}{r_{Fe}^2} = \frac{3}{2}kT
\label{balance}
\end{equation}
The radius of the iron atomic complex paramagnetically bound
to the monopole at $T\approx1200^0~C$ is:
\begin{equation}
r_{Fe} \simeq 6 \cdot 10^{-8} cm.
\label{diam}
\end{equation}

Such complexes contains about  30  atoms  of iron.
Considering, that  the  movement  of  such  small  blob  is
determined by the Stokes law:
\begin{equation}
F_v = 6\pi \cdot r_{Fe}\cdot \eta  v~,
\label{stocs}
\end{equation}
where $v$  is  the  monopole  velocity,  $\eta$  is  the  dynamic
viscosity of liquid foundry iron,
$\eta = 2\cdot 10^{-3} kg/(m \cdot sec)$ at $T = 1250^0~C$.

Equating the force of friction to the gravity $F_g = mg$, we find
the velocity $v$ of the monopole falling through the melt
\begin{equation}
v = \frac{mg}{6\pi \cdot r_{Fe} \cdot \eta}~,
\label{radius}
\end{equation}
that makes $v \simeq 3\cdot 10^{-1} m/sec$ for  a  monopole  with
mass about $10^{15}~GeV$.

This corresponds to kinetic energy of  the  complex  about  $10^6
~eV$, which is large enough for a skinning off of the complex  at
the solid bottom of the furnace. Let's remark, that from a formal
point such an approach  is  quite  acceptable,  as  the  Reynolds
number in our case it is not large enough: $Re<10^{-3}$.

This grain (complex) should sink in the  liquid  iron  at  10-100
cm/s velocity until it reaches the bottom of the  blast  furnace.
Then the atoms of iron are  stripped  off  the  monopole  in  the
material of the oven bottom,  and  the  monopole  falls  further,
accelerating up to the velocity of sound in the matter.

We should have in  mind,  that the energy loss of monopole due
to the Cherenkov radiation of phonons and magnons in a medium
reach $10^8~eV/cm $ and  even  more, as it was shown  earlier
\cite{VK},\cite{e-print}. This is a main point to understand,
that  the gravitation forces  cannot  accelerate  the
monopole  up to velocity higher than the speed of sound in a medium.

Usually the blast furnace melts about 10 000 tons of ore per  day,
and it is possible without great problems to expose the MO-detector,
for example, during one year. Thus we hope that such  MO-detector
can improve significantly the experimental limit on  the  density
of relict monopoles entrapped in iron ore, which today  is  equal
to $\rho_m < 2\cdot 10^{-7}/g$~\cite{PDG}.

Furthermore, in a sinter machine the ore is also heated above the
Curie temperature, but not up to the melting point. So, to  shake
off the iron atoms, we have to kick  the  iron  ore  pieces  with
acceleration of $10-100~g.$ Clearly, in this last case
the probability of monopole release is considerably lower.

\section{Conclusion.}

The interaction of monopoles with films of magnetic materials  is
considered. In particular, the interaction of slow monopoles with
thin films of easy-axis magnetics with high and low  mobility  of
domain walls (materials with magnetic bubbles)is discussed. It is
shown, that during the movement of a slow  monopole  through  the
magneto-hard magnetic film, a track-domain  can  be  formed  with
typical size of about  50~nm  and  with  magnetization  of  about
several thousand Oe. Thus the magnetic flux of the track  appears
to be about the value of the flux quantum. For detection of  such
a flux, the detectors using fluxmeter on  the  basis  of  already
widely known SQUID can be used.

It appears  that  for  registration  of  traces  of  slow  cosmic
monopoles in magnetic matter, the experimental devices using  the
Kerr magneto-optical  effect  are  more  appropriate.  They  have
emerged recently in a shape suitable for detector design, with an
appropriate adjustment.

It should  be  realized  that  such  passive  detectors  will  be
especially effective in search of relict monopoles, entrapped  in
ferromagnetic inclusions of iron ore. These monopoles  should  be
extracted from the ore during melting process.  Then  these  slow
moving monopoles can be detected by a passive MO detector. We can
expose MO disks in a cavity under a blast furnace  exactly  under
the bath with melting  metal,  where  the  temperature  does  not
exceed $+50^0~C$. In the melting process the temperature  of  ore
exceeds  the  Curie  point  and  its  ferromagnetic    properties
disappear. Hence the ferromagnetic traps which hold the monopoles
are "switched off" and the released monopoles  fall  through  the
melting metal to the bottom of the bath and finally  through  the
MO disks. While a monopole, moved downward by the gravity  force
crosses the surface of one of the MO disks, it leaves a  magnetic
track in its coat. It is  possible  to  obtain  the  slow  moving
relict monopole also by the  sinter  machine,  and  usually  both
installations process no less than 10 000 tons of ore per day.

The authors are grateful to all colleagues for helpful and lively
discussions of  this  work  in  various  places  and  institutes.
Special thanks go for stimulating interest and useful remarks to
L.M.~Barkov, I.B.~Khriplovich (all from Budker INP)
and V.A. Gordeev (PNPI).


\begin{thebibliography}{99}
\bibitem{Dir} P.A.M. ~Dirac, Proc. Roy. Soc. {\bf A133} (1931) 60.
\bibitem{Groom} D.E.~Groom, Phys. Rep. {\bf 140} (1986).
\bibitem{Klapdor}  H.V.~Klapdor-Kleingrothaus, A.~Staudt,
{\it Teilchenphysik ohne Beschleuniger}, Stuttgart, 1995
\bibitem{Barkov} L.M.~Barkov,I.I.~Gurevich, M.S.~Zolotarev et al.,
Zh. Eksp. Teor. Fiz {\bf 61} (1971) 1721, [JETP 61 (1971) 1721]
\bibitem{D0} D0 Collaboration, S.~Abachi et al., preprint Fermilab
             Pub-98/095-E,(1998).
\bibitem{Pol} A.M.~Poljakov, Pis'ma Zh. Eksp. Teor. Fiz. {\bf 20}
(1974) 430.\\ JETP Lett. 20 (1974) 194.
\bibitem{tHooft} G.~'t~Hooft, Nucl.Phys. B {\bf 79}, 276 (1974).
\bibitem{Mono} {\it Dirac monopole}. Ed. B.M.~Boltovsky  and
               Yu.D.~Usachev,"Mir", 1970.
\bibitem{Cabr} B.~Cabrera, Phys. Rev. Lett. {\bf 48} (1982) 1378.\\
               R.D.~Gardner et al. Phys. Rev. {\bf 44} (1991) 622.
\bibitem{MACRO} MACRO collaboration, Phys. Lett. {\bf B406} (1997) 249.
\bibitem{Adams} F.C.~Adams et al., Phys. Rev. Lett. {\bf 70} (1993) 2511.
\bibitem{limit} B.C.~Choudary for MACRO Collaboration, LANL  e-print
archive hep-ex/9905023.
\bibitem{macro_l}  Ambrosio, M. et al.,  MACRO Coll., 1999, hep-ex/9904031,
EPJ C ( in press)\\
\bibitem{Rub} V.A.~Rubakov, Pis'ma v ZhETF {\bf 33}(1981)644,\\
              Nucl. Phys. {\bf B203} (1982) 311
\bibitem{Cal} C.~Callan, Phys. Rev. D {\bf 25} (1982) 2141.
\bibitem{Belo} I.A.~Belolaptikov et  al.  LANL  e-print  archive
               astro-ph/9802223.
\bibitem{MO} Ed.: Glenn T. Sincerbox, James M. Zavislan,
    {\it (Selected Papers on Optical Storage)}, Vol. MS 49, SPIE - The
    International Society for Optical Engineering, 1992.
\bibitem{CR39} B. Ichinose et al. Nucl. Inst. Meth.{\bf A286}(1989)327.\\
    MACRO Collaboration, S. Ahlen et al.,~ Search for magnetic  monopoles
    with MACRO track etch detector",~ LNGS 94/115 (1994).
\bibitem{VK} P.V.~Vorob'ev, I.V. ~Kolokolov. Preprint BINP 98-16,
             Novosibirsk 1998, LANL E-print hep-ph/9806495,\\
       P.V.~Vorob'ev, I.V. ~Kolokolov. Pis'ma Zh. Eksp. Teor. Fiz.
      {\bf 67} (1998) 866.
\bibitem{e-print}I.V. ~Kolokolov, P.V.~Vorob'ev, V.V.~Ianovski.
                 LANL E-print hep-ph/9809420;\\
      In:  Proceedings of the XXXII PNPI Winter School.
      (Particle and Nuclear Physics). Ed. J.Azimov, V. Bunakov, V.
      Gordeev. SPb, 1998, p.134-152.
\bibitem{Mopt} Chung-Hee Chang, M.H.Kryder.
        J.Appl. Phys. {\bf 75} (1994) 142.
\bibitem{Bell} (http://
    portal.research.bell-labs.com/leisure/souvenirs/gallery/bits.html)
\bibitem{PDG} Review of Particle Physics. Evro.Phys.J. 3 (1998) 742,
              PRD 54 (1996) 686.
\end{thebibliography}
\end{document}